\documentclass[twocolumn]{aa} 
\usepackage{graphicx}
\usepackage{txfonts}
\begin{document}
   \title{ 
Erratum: Detection of diffuse TeV gamma-ray emission from the
nearby starburst galaxy NGC 253
}
   \titlerunning{Erratum: TeV gamma-ray from NGC 253}


\authorrunning{Itoh, Enomoto, Yanagita, Yoshida et al.}
   \author{C.~Itoh\inst{1}\and
R.~Enomoto\inst{2}\thanks{\email{enomoto@icrr.u-tokyo.ac.jp}}\and
S.~Yanagita \inst{1}\and T.~Yoshida\inst{1}\and
A.~Asahara \inst{3}\and
G.V.~Bicknell\inst{4}\and
R.W.~Clay\inst{5}\and
P.G.~Edwards\inst{6}\and
S.~Gunji\inst{7}\and
S.~Hara\inst{3,8}\and
T.~Hara\inst{9}\and
T.~Hattori\inst{10}\and
Shin.~Hayashi\inst{11}\and
Sei.~Hayashi\inst{11}\and
S.~Kabuki\inst{2}\and
F.~Kajino\inst{11}\and
H.~Katagiri\inst{2}\and
A.~Kawachi\inst{2}\and
T.~Kifune\inst{12}\and
H.~Kubo \inst{3}\and
J.~Kushida\inst{3,8}\and
Y.~Matsubara\inst{13}\and
Y.~Mizumoto\inst{14}\and
M.~Mori\inst{2}\and
H.~Moro\inst{10}\and
H.~Muraishi\inst{15}\and
Y.~Muraki\inst{13}\and
T.~Naito\inst{9}\and
T.~Nakase\inst{10}\and
D.~Nishida\inst{3}\and
K.~Nishijima\inst{10}\and
K.~Okumura \inst{2}\and
M.~Ohishi\inst{2}\and
J.R.~Patterson\inst{5}\and
R.J.~Protheroe\inst{5}\and
K.~Sakurazawa\inst{8}\and
D.L.~Swaby\inst{5}\and
T.~Tanimori \inst{3}\and
F.~Tokanai\inst{7}\and
K.~Tsuchiya\inst{2}\and
H.~Tsunoo\inst{2}\and
T.~Uchida\inst{2}\and
A.~Watanabe \inst{7}\and
S.~Watanabe\inst{3}\and
T.~Yoshikoshi \inst{16}}

   \offprints{R. Enomoto}

   \institute{
Faculty of Science, Ibaraki University,
Mito, Ibaraki 310-8512, Japan
\and
Institute for Cosmic Ray Research, Univ. of Tokyo, Kashiwa,
Chiba 277-8582, Japan
\and
Department of Physics, Kyoto University, Sakyo-ku, Kyoto 606-8502, Japan
\and
MSSSO, Australian National University, ACT 2611, Australia
\and
Department of Physics and Math. Physics, University of Adelaide, SA 5005,
Australia
\and
Institute for Space and Aeronautical Science, Sagamihara,
Kanagawa 229-8510, Japan
\and
Department of Physics, Yamagata University, Yamagata, Yamagata 990-8560, Japan
\and
Department of Physics, Tokyo Institute of Technology, Meguro-ku, Tokyo 152-8551, Japan
\and
Faculty of Management Information, Yamanashi Gakuin University,
Kofu,Yamanashi 400-8575, Japan
\and
Department of Physics, Tokai University, Hiratsuka, Kanagawa 259-1292, Japan
\and
Department of Physics, Konan University,
Hyogo 658-8501, Japan
\and
Faculty of Engineering, Shinshu University, Nagano, Nagano 380-8553, Japan
\and
STE Laboratory, Nagoya University, Nagoya, Aichi 464-8601, Japan
\and
National Astronomical Observatory of Japan, Mitaka, Tokyo 181-8588, Japan
\and
Department of Radiological Sciences,
Ibaraki Prefectural University of Health Sciences,
Ibaraki 300-0394, Japan
\and
Department of Physics, Osaka City University, Osaka, Osaka 558-8585, Japan}

   \date{A\&A {\bf 396}, L1-L4(2002). A\&A {\bf 402}, 443-455(2003)}

 
  \abstract
   {
   }
   {
The CANGAROO-II telescope observed sub-TeV gamma-ray emission from the 
nearby starburst galaxy NGC 253.
The emission region was extended with a radial size of 0.3-0.6 degree. 
On the contrary, H.E.S.S could not confirm
this emission and gave upper limits at the level of the CANGAROO-II 
flux. In order to resolve this discrepancy, we
analyzed new observational results for NGC 253 by CANGAROO-III and 
also assessed the results by CANGAROO-II.
 }
   {
Observation was made with three telescopes of the CANGAROO-III in 
October 2004. We analyzed three-fold coincidence
data by the robust Fisher Discriminant method to discriminate gamma 
ray events from hadron events.
   }
   {
\begin{it}
The result by the CANGAROO-III was negative. 
\end{it}
The upper limit of gamma 
ray flux was 5.8\% Crab at 0.58 TeV for point-source assumption.
In addition, the significance of the excess flux of 
gamma-rays by the CANGAROO-II 
was lowered to less than 4 sigma after
assessing treatment of malfunction of photomultiplier  tubes.
   }
   {}

   \keywords{
gamma rays: observation -- galaxies: starburst -- galaxies: 
individual: NGC 253 -- galaxies: halos: cosmic rays
               }

   \maketitle
%

\section{Introduction}

NGC~253 is a nearby ($d=2.5$\,Mpc) 
(de Vaucouleurs, \cite{Vaucouleurs}), normal spiral,
starburst, and edge-on galaxy.  
Starburst galaxies are generally expected to have
cosmic-ray energy densities about hundred times larger than that of our
Galaxy (Voelk et al., \cite{Voelk}) due to the high 
rates of massive star formation
and supernova explosions in their nuclear regions. The star-formation
rates can be estimated from the far-infrared (FIR) luminosities,
and the supernova rates can be also inferred based on the assumption of an
initial mass function. Since the supernova rate of NGC~253 is
estimated to be about 0.05 - 0.2 yr$^{-1}$ 
(Mattila and Meikle, \cite{Mattila}, Antonucci and Ulvestad, 
\cite{Antonucci}, van Buren and Greenhouse, \cite{Buren}), a high cosmic-ray
production rate is expected in this galaxy.

Although there were no non-thermal X-ray nor GeV gamma-rays detections yet,
an extended synchrotron-emitting halo
of relativistic electrons was observed (Carilli et al., \cite{Carilli}). 
The halo extends to a large-scale height, 
where inverse Compton scattering (ICS) may be a more
important process for gamma-ray production than pion decay and
bremsstrahlung. The seed photons for ICS are expected to be mainly FIR
photons up to a few kpc from the nucleus, and cosmic microwave
background radiation at larger distances.

In 2002 CANGAROO-II reported on the detection of a diffuse TeV gamma-rays 
in the direction of NGC~253 (Itoh et al. \cite{Itoh2002,itoh_aa}).
The estimated size was 0.3 $\sim$ 0.6 degrees in radius.
The emission was later interpreted as halo-like (Itoh et al., \cite{itoh_apj}).
H.E.S.S., however, claimed null results on them (Aharonian et al. \cite{hess}).  
The upper limits were located in marginal values 
(actually H.E.S.S.'s upper limits
crossed over with CANGAROO-II's fluxes around TeV).
The main purpose of this report is to clarify this.
H.E.S.S. also discussed calorimetric gamma-ray emission at the very central
region of this galaxy in that report. The point source search at the center
of this galaxy, therefore,
is also subjected.

In this purpose,
we observed NGC~253 with the CANGAROO-III telescope in 2004 October.
In this paper we describe results of this observation with three telescopes
coincidence.
The responsibility of this part (Section 2-4) is taken by
the authors of Enomoto et al. (\cite{enomoto_0852}).
Also the discussion on the previous CANGAROO-II analysis is included.


\section{Observation}

CANGAROO-III is one of two major imaging atmospheric
Cherenkov telescopes located in the southern
hemisphere.
The CANGAROO-III stereoscopic system consists of four imaging atmospheric
imaging telescopes located near Woomera, South Australia (31$^\circ$S,
137$^\circ$E).
Each telescope has a 10-m$\phi$ reflector.
Each reflector consists of 114 segmented spherical mirrors (80\,cm
in diameter with a radius of curvature of 16.4\,m) made of FRP
(Kawachi et al. \cite{kawachi}) mounted on a parabolic
frame ($f/d$=0.77, i.e., a focal length of 8\,m).
The total light collection area is 57.3\,m$^2$.
The first telescope, T1, which was the CANGAROO-II telescope
(Itoh et al. \cite{itoh_aa}),
is not presently in use due to its smaller field of view
and higher energy threshold.
The second, third, and fourth telescopes (T2, T3, and T4) were used for the
observations described here.
The camera systems for T2, T3, and T4 are identical and their details
are given in Kabuki et al.(\cite{kabuki}).
The telescopes are located at the 
east (T1), west (T2), south (T3) and north (T4)
corners of a diamond 
with sides of $\sim$100\,m (Enomoto et al. \cite{enomoto_app}).

The observations were carried out 
in the period from 2004 October 7 to 17
using ``wobble mode"
in which the pointing position of each telescope was
shifted in declination between $\pm$0.5 degree from
the center of the galaxy (RA, dec = 11.888$^\circ$, $-$25.288$^\circ$
J2000) 
every 20 minutes (Daum et al. \cite{wobble}).
Data were recorded for T2, T3 and T4 when
more than four photomultiplier (PMT) signals 
exceeded 7.6 photoelectrons (p.e.) 
in any telescope.
The GPS time stamp was recorded in each telescope dataset. 
An offline coincidence of time stamps within
$\pm$100\,$\mu$s (Enomoto et al. \cite{enomoto_vela}) 
was required for a stereo event.
The typical trigger rate for each telescope was 80\,Hz, which was
reduced to 10\,Hz for stereo events
for three-fold coincidence.
Each night was divided into two or three periods, i.e., ON--OFF,
OFF--ON--OFF, or OFF--ON observations. 
Note that the OFF-source observations were also made in
''wobble mode".
This is carried out because the previously we claimed the detection of 
the diffuse source.
ON-source observations were timed
to contain the meridian passage of the target. 
On average the OFF source regions were located with an offset in RA of 
+30$^\circ$ or $-$30$^\circ$ from
the center of the galaxy. 
The total observation time was 1179 and 753~min, for ON and OFF 
observations, respectively.

Next we required the images in all three telescopes to have clusters
of at least five adjacent pixels exceeding a 5\,p.e.\ threshold
(three-fold coincidence).
The event rate was reduced to $\sim$6\,Hz by this criterion.
Looking at the time dependence of these rates, we can remove data
taken in cloudy
conditions. This procedure is the same as the ``cloud cut''
used in the CANGAROO-II analysis (Enomoto et al. \cite{enomoto_nature}).
We also rejected data taken at elevation angles less than 70$^\circ$.
In total, 750 min.\ data survived these cuts for ON and
517 min.\ for OFF, with a
mean elevation angle of 78.6$^\circ$.

The light collecting efficiencies, including the reflectivity
of the segmented mirrors, the light guides, and the quantum efficiencies
of photomultiplier tubes were monitored by a muon-ring analysis
(Enomoto et al. \cite{enomoto_vela}). 
The light yield per unit arc-length is approximately proportional
to the light collecting efficiencies.
The ratios of these at the observation period with respect to the
mirror production times (i.e., deterioration factors) 
were estimated to be 45, 55, and 73\% for
T2, T3, and T4, respectively. The measurement errors are considered to
be at less than the 5\% level.
These values were checked analyzing Crab data which were obtained in
2004 November described in Enomoto et al.(\cite{enomoto_0852}).
The deteriorations were mostly due to dirt and dust settling on the
mirrors.
We cleaned the mirrors with water in October 2005 
and the partial improvement (a factor of 1.3-1.4) of
the light collecting efficiencies were observed.


\section{Analysis}

The analysis procedures used were identical with those 
described in Enomoto et al. (\cite{enomoto_vela,enomoto_0852}),
we omit a detailed discussion here.

At first, the Hillas parameters (Hillas \cite{hillas}) 
were calculated for the three telescopes' images.
The gamma-ray incidence directions were adjusted by minimizing
the sum of squared widths (weighted by the photon yield) 
of the three images seen from the assumed position (fitting parameter).
Then the Fisher Discriminant (hereafter $FD$ in short) (Fisher \cite{fisher})
is calculated.
The input parameters are
energy corrected $widths$ and $lengths$ for the
T2, T3, and T4.

Since we have $FD$ distributions for OFF-source data and the 
Monte-Carlo gamma-ray
events, we can assume these are background and signal behaviors.
We, therefore, can fit the $FD$ distribution of ON
with the above emulated
signal and real background functions, to derive the number of signal events
passing the selection criteria. 
With this fit, we can determine the gamma-ray excess without any positional
subtractions, i.e., appropriate for diffuse radiations.
This is a two-parameter fitting and these coefficients can be exactly
derived analytically.

This method was checked by analysis of Crab nebula data taken in November
2004. The wobble-mode observation was also used. 
The analyzable data corresponded to 316.4 min. 
The flux is 1.2$\pm$0.3 times the standard Crab flux with the power-law
consistent with the standard index of $-$2.5.


\section{Results}

\begin{figure}
   \centering
   \includegraphics[width=8cm]{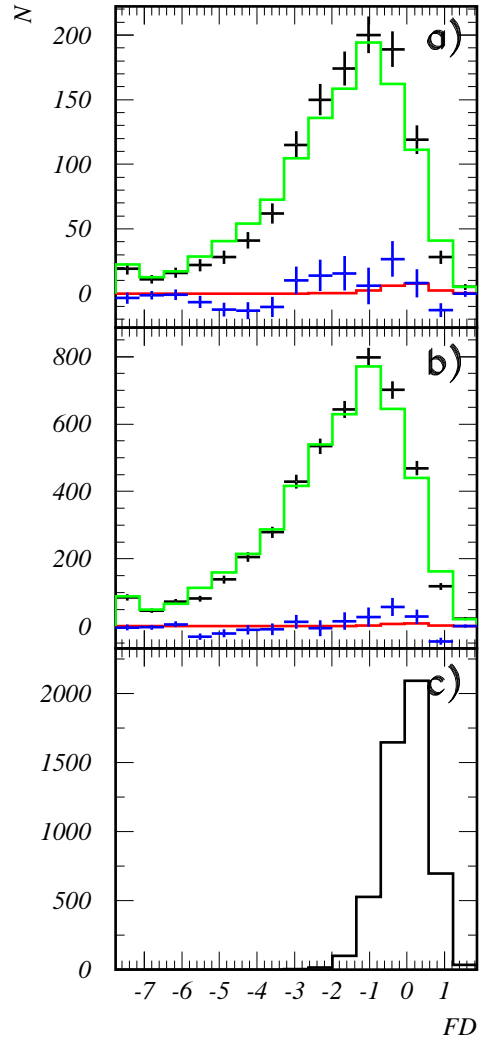}
      \caption{
Fisher Discriminant ($FD$) distributions;
a) for $\theta^2~<~0.05$ degree$^2$ (point-source assumption),
b) $\theta^2~<~0.25$ degree$^2$ (0.5-degree diffuse),
and c) the Monte-Carlo gamma-ray events.
The black data point with error bars were obtained from
the ON source runs. The green histograms were made by
the OFF source runs.
Note that the vertical normalization of each histogram was a result of
the fitting procedure described in the text.
The blue points were the background-subtracted data and the red
histograms are best fitted signals.
              }
         \label{ffd}
   \end{figure}

The signal function for $FD$ is shown by the black histogram in 
Fig. \ref{ffd}-c).
That for background was made from the region $\theta^2~<$ 0.5 degree$^2$
in the OFF data (the green histograms in Fig. \ref{ffd}-a) and b)).
As has been described in the previous section, we carried out
two-parameters fit, one is the vertical normalization of the background
shape and the other is of the signal.
The best fit results are shown in Fig. \ref{ffd} a) and b).
The black data points with error bars are ON data.
The background subtracted signals are shown by the blue points.
The red histograms are the best fitted yield for signals.
The entry within $\theta^2~<$ 0.05 degree$^2$ is plotted in Fig. \ref{ffd}-a)
and b) is for $\theta^2~<$ 0.25 degree$^2$, i.e., a) for the point source
assumption and b) for 0.5 degree diffuse assumption, respectively.
In the both regions, we did not see any statistically significant signals.
The threshold of this analysis was estimated to be 0.58 TeV.

\begin{figure}
   \centering
   \includegraphics[width=8cm]{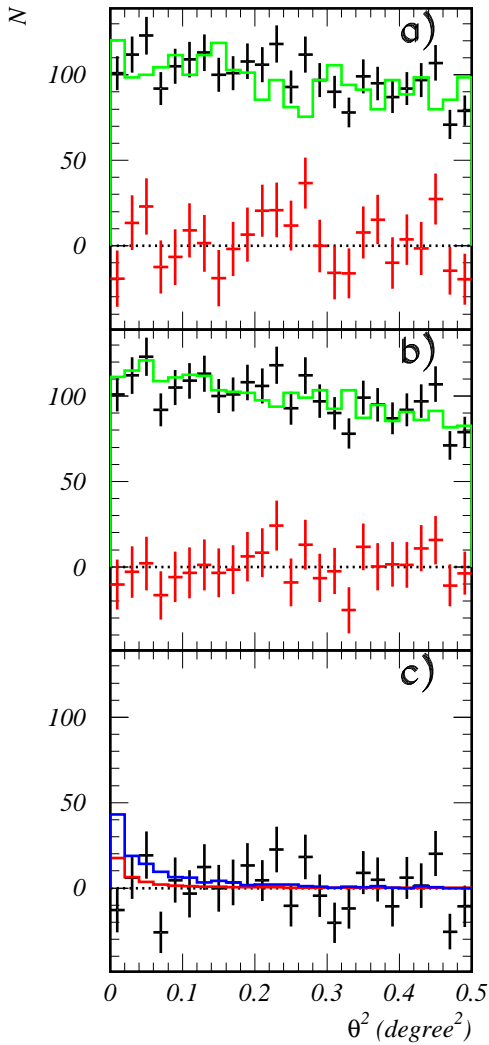}
      \caption{
$\theta^2$ (degree$^2$) distributions
for $|FD|~<$ 1, a) and b).
The black points with error bars are obtained by the ON
source runs. The green histogram in a) was obtained
by the OFF source runs which was normalized by the observation
time.
The green histogram in b) was by ''wobble background" 
from ON source runs
normalized by the inverse of the number of
''wobble points" (= 1/6).
The red points were obtained by the background subtractions.
Zero level is shown by the dotted line.
The black points in c) is obtained by the fitting procedure described
in the text.
The red histogram in c) is 2$\sigma$ upper limit 
for the point source assumption.
The blue in c) is the expected yield assuming the CANGAROO-II flux.
              }
         \label{ftheta2}
   \end{figure}

Then we study spatial distributions of gamma-ray like events.
At first we select these by $|FD| ~<$ 1 ( see Fig \ref{ftheta2}-c).
The black points with error bars in a) and b) are the ON data. The green
histogram in a) is the OFF data with the normalization based on 
ON/OFF observation times.
These two agree well, i.e., there is no signal anywhere in this plotting
range.
The red points in a) and b) are the background subtracted data.
Since the statistics of the OFF-source run is limited, the errors
in the background-subtracted data are dominated by this.

For the point source assumption, we can use "wobble"-background analysis.
The signal region for it is $\theta^2~<$ 0.05 degree$^2$, therefore,
we obtain six background points. The sum of them with a normalization
factor of 1/6 is shown by the green histogram in Fig. \ref{ftheta2}-b).
Now the error due to the subtraction becomes small, however, we again can not
see any signal excess.

We made $FD$ distributions for $\theta^2$ slices and carried out the same
fitting procedure as in the case of Figs \ref{ffd}.
The excesses obtained are plotted in Fig. \ref{ftheta2}-c) (the
black points).
The red histogram is a 2$\sigma$ upper limit (37.5 events) for signal for
point-source assumption.
Actually $\chi^2$-minimum of this fit has negative excess. We, therefore,
constrained that the excess is positive in deriving upper limit.
The result is 5.8\% Crab at 0.58 TeV, 
a factor worse than H.E.S.S.'s upper limit.
For references, the upper limits obtained from the red histograms in
Fig. \ref{ftheta2}-a) and b) are 50 and 38 events, respectively.
The blue is an expected yield under the assumption of the CANGAROO-II flux
(Itoh et al. \cite{itoh_aa}) for point source assumption.
Our upper limit is inconsistent with the CANGAROO-II detection under
the assumption of the point-source and is marginal under that of
0.5-degree diffuse emission.

\begin{figure}
   \centering
   \includegraphics[width=8cm]{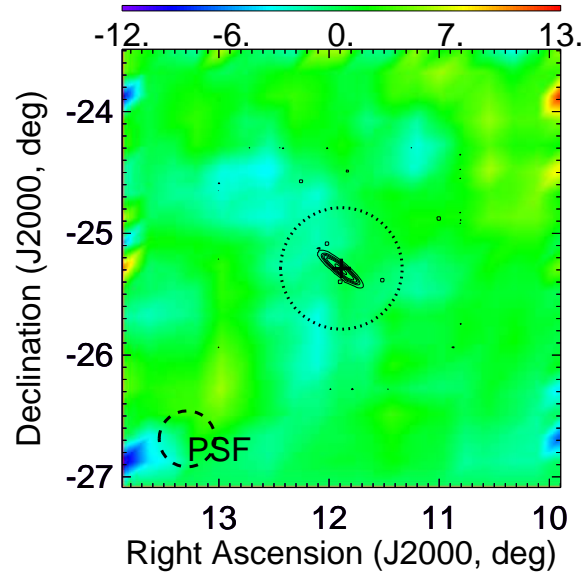}
      \caption{
Excess count map.
The rainbow map is the excess count. The black contour is DSS2
(second version of Digital Sky Survey) data.
The dotted circle is 0.5 degree radius.
The point spread function is shown in left-below corner
(the dashed line).
              }
         \label{fmap}
   \end{figure}

We also searched for signals in the broad range of such as 3.8 $\times$
3.8 degree$^2$. In total 316 $FD$ distributions were 
made by the spatial bin size
of 0.2 $\times$ 0.2 degree$^2$.
The signal function is the same as in Fig. \ref{ffd}-c). Each background
function was made from the OFF-source runs with the bin size
of 0.6 $\times$ 0.6 degree$^2$ with the same center position as ON
data points in order to compensate the background
statistical level.
The number of gamma-ray like events are shown in Fig. \ref{fmap}.
The contours are DSS2 data. The 0.5-degree circle is shown by the dotted line.
The point spread function is shown in the left-lower corner.
The color scale of the excess counts is shown on top.
The region within 0.5 degree circle is zero consistent, also
even in the surrounding region, we can not find any excess.

\begin{figure}
   \centering
   \includegraphics[width=8cm]{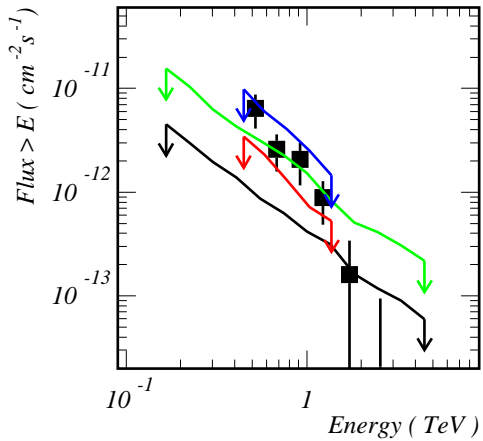}
      \caption{
Integral fluxes.
The points with error bars are the CANGAROO-II's ones
(see text for the detail).
The black curve is 99\% upper limit (UL) by H.E.S.S. for
point source assumption. The green is that for 0.5 degree
diffuse source.
The red is 2$\sigma$ UL for this observation for point source
assumption and the blue for 0.5 degree diffuse.
              }
         \label{fflux}
   \end{figure}

The same kind of analysis was repeated with five different
energy thresholds
estimated from the total number of photoelectrons. 
Here, H.E.S.S., showed integral flux upper limits and we follow
it in order to compare with them.
The upper limit of the integral flux versus energy
is obtained and is shown in Fig. \ref{fflux}.
The red line is 2$\sigma$ upper limit for the point-source assumption.
The blue is that for 0.5-degree diffuse case.
The CANGAROO-II data points were obtained from Table 6 
of Itoh et al. (\cite{itoh_aa}). Since they are differential, we multiplied
$E/(\gamma-1)$ bin by bin bases (the black points with error bars). 
The power-law index $\gamma$ was assumed
to be 3.85 which was the best fitted value in the same reference.
They are slightly harder than those in Fig. 2. of 
Aharonian et al. (\cite{hess}),  which we do not know why there is difference.
The black line is their upper limit for the point-source assumption
and the green that for 0.5-degree diffuse.


\section{Discussion}

Our upper limits are 2$\sim$3 times higher than 
those obtained by H.E.S.S. (Aharonian et al.
\cite{hess}).
These factors can be understood by the blur spot sizes of
the segmented mirrors (0.14, 0.12, and 0.09 degrees for T2, T3, and T4,
respectively).
The effective area of three telescopes (when three-fold coincidence is 
required) is
smaller than that by a single telescope measurement,
i.e., the threshold is higher.
Also important thing is that 
the CANGAROO-II carried out multiple-years ( and multiple-months per
year ) observation, while this is single year and single month one
(or rather slightly larger than a single week) observation.
Therefore, the previous CANGAROO-II
fluxes (Itoh et al. \cite{itoh_aa}) are located between our upper limits
under two assumptions of the point and diffuse sources.
CANGAROO-III data are not fully denying the CANGAROO-II result.
We need to accumulate several times (4 $\sim$ 9) more observation time
in order to reach H.E.S.S.'s level when we assume that there is
no signal. 

Before doing it, we need to consider the fact that 
the CANGAROO-II data and its analysis
software are still available.
We checked the previous CANGAROO-II analysis in detail. The
detailed description can be found in Itoh et al. (\cite{itoh_aa}).
We found an improper part
which can be found in the description written in 
Section 3.6 in Itoh et al. (\cite{itoh_aa}), that is
the procedure to remove hot channel.
In the previous CANGAROO-II analysis, 
the deformation of $\alpha$ spectrum appeared in OFF data
(non-flat $\alpha$ distribution)
(image oriented angle: Hillas parameter (Hillas \cite{hillas})).
Generally hot pixels deform $\alpha$ spectrum.
We carried out the following procedure to find those bad pixels;
\begin{itemize}
  \item 
  Hot ``box" scan for recovering flatness were carried out,
  where ``box" is a unit of sixteen (four by four) 
  neighbored photo-multiplier tubes ( as shown in 
  Fig. 1 of Itoh et al. \cite{itoh_aa}).
  \item further scan inside these sixteen channel were done
  and finally find the field-deforming pixels.
\end{itemize}
Note that this was not applied to
RX J1713.7-3946 (Enomoto et al. \cite{enomoto_nature})
, Galactic Center (Tsuchiya et al. \cite{tsuchiya}), 
nor RX J0852.0-4622 (Katagiri et al. \cite{katagiri}).
For RX J1713.7-3946, we removed
hot pixels due to small discharges triggered by the bright star passages.
For Galactic center and
RX J0852.0-4622, we selected them
based on the $\chi^2$ calculated by the pixel-hit rate and deviation
of each ADC spectrum from the average one.
These three observations had bright stars in the field of view (FOV).
On the other hand, the FOV of the NGC 253 observation did not
contained any bright ones, i.e., it was relatively dark field.
Although there were no explicit high hit-rate pixels, the deformation
of $\alpha$ spectrum appeared in OFF data.
This is why we adopted the above procedure.
These rejections for masked pixels were applied commonly to
the ON and OFF runs.
We, therefore, thought it was unbiased.
We, however, found that there is a big discrepancy of the excess events
before and after this procedure.
Their numbers were
700 and 2000, respectively.

Excess of 2000 events which was 11$\sigma$ is now reduced to be 
less than 4$\sigma$,
that is lower than the standard of claiming a positive signal.
Assuming 2$\sigma$ upper limit, it is now clear that at most
a half level of signal is allowed compared to the previous flux level
(Itoh et al. \cite{itoh_aa}),
which is lower than the upper limit by H.E.S.S. of extended source assumption.
In this case, the new expected yield for this observation would be
approximately the red histogram in Fig. \ref{ftheta2}-c)
in the point-source assumption.

To summarize the present situation, 
we have nothing to deny H.E.S.S.'s observation, i.e.
for diffuse radiation of order 0.5 degree
the emission should be less than 6\% Crab and for the point-source
it is less than 2\% Crab at 300 GeV.
However, the 
physics interest on this astronomical object is not lost.
In fact, H.E.S.S. discussed the possibility of calorimetric 
gamma-ray emission in the starburst region (Aharonian et al. \cite{hess}).
Also the radio halo should be originated by the high energy electrons (Carilli
et al. \cite{Carilli}). 
The fine resolution (spatial and energy) and high sensitivity 
with also wide energy range
observation is still awaited both for the point and diffuse sources.


\section{Conclusions}

We have observed the nearby starburst galaxy NGC 253 in October 2004.
TeV gamma-rays were searched for in the data obtained by three
telescopes. No statistically significant signals were obtained
for both assumptions of point and diffuse source.
Our upper limits were marginally inconsistent with the previous
CANGAROO-II observation. We, therefore, further investigated the
previous analysis and found an improper procedure in hot channel
rejection algorithm. After removing that procedure, the previous
CANGAROO-II flux was reduced less than a half.
We concluded that we can not claim 
any evidence for gamma-ray emission from NGC 253.

\begin{acknowledgements}

We thank Dr. Jim Hinton for providing the upper limits value of H.E.S.S.
This work was supported by a Grant-in-Aid for Scientific Research by
the Japan Ministry of Education, Culture, Sports, Science and Technology, 
the Australian Research Council, JSPS Research Fellowships,
and Inter-University Researches Program 
by the Institute for Cosmic Ray Research.
We thank the Defense Support Center Woomera and BAE Systems.

\end{acknowledgements}

\end{document}